\documentclass[aps,prb,letterpaper,amsmath,amssymb,superscriptaddress,footinbib,twocolumn]{revtex4-2}

\usepackage{graphicx}
\usepackage[usenames]{color}
\usepackage[colorlinks=true,linkcolor=magenta,citecolor=magenta,anchorcolor=green,urlcolor=magenta]{hyperref}
\usepackage{mathrsfs}

\usepackage[percent]{overpic}
\usepackage{tabularx}
\usepackage{multirow}
\usepackage{orcidlink}

\usepackage{CJK}

\def\be{\begin{equation}}
\def\ee{\end{equation}}
\def\ba{\begin{aligned}}
\def\ea{\end{aligned}}
\def\bea{\begin{eqnarray}}
\def\eea{\end{eqnarray}}

\begin{document}
\begin{CJK*}{UTF8}{}
\title{Universal meson spectra near $(1+1)$-dimensional Ising criticality}
\author{Xiao Wang \CJKfamily{gbsn}(王骁)~\orcidlink{0000-0003-2898-3355}}
\affiliation{Department of Physics, Cornell University, Ithaca, NY, USA}

\author{Jianda Wu \CJKfamily{gbsn}(吴建达)~\orcidlink{0000-0002-3571-3348}}
\altaffiliation{wujd@tongji.edu.cn}
\affiliation{School of Physics Science and Engineering, Tongji University, Shanghai  200092, P. R. China}

\begin{abstract}
Near $(1+1)$-dimensional [$(1+1)$D] Ising criticality,
a magnetic perturbation induces confinement and produces a cascade of bound-state excitations known as mesons.
Here we show these mesons share a universal mass scaling after independently 
rescaling the model-dependent microscopic couplings.
The number of stable mesons is controlled by the lightest two-meson threshold,
while the lightest-meson mass follows a continuous trajectory
characterized by a single scaling parameter.
Using Hamiltonian truncation method, 
we obtain the trajectory numerically in both Ising field theory and the near-critical mixed-field Ising chain (MFIC).
Under the rescaling,
the trajectory and stable-meson-count crossover windows of MFIC both collapse onto the field-theory results.
To further demonstrate the above universal organization of the meson spectra,
we consider a class of four-periodic spin-$1/2$ Heisenberg-Ising chains under transverse fields,
whose parameter space contains a family of quantum Ising critical points.
The Hamiltonian family includes effective spin models for the quasi-one-dimensional antiferromagnets Ba(Sr)Co$_2$V$_2$O$_8$.
Using tensor-network calculations,
we obtain the corresponding lightest-meson mass trajectory for BaCo$_2$V$_2$O$_8$ and find that it also collapse onto the same universal curve given by field-theory result.
Our results suggest that the universal scaling structure of quantum Ising criticality extends into the nearby confining regime,
governing the organization of the meson spectrum.
They thereby provide a practical criterion for interpreting excitations of quasi-1D Ising-like magnets in mixed fields
beyond $E_8$ integrability.
\end{abstract}

\maketitle
\end{CJK*}

\textit{Introduction.}---
Universality has been a central organizing principle in modern physics since the advent of scaling theory and the renormalization group more than half a century ago
\cite{leo_1966,alexander_1970,kenneth_1971_1,kenneth_1971_2,alexander_1984}.
Near a continuous phase transition, 
long-range correlation is governed by an infrared (IR) fixed point and is 
therefore insensitive to most microscopic ultraviolet (UV) details.
In one spatial dimension, 
many quantum critical universality classes are described by $(1+1)$ dimension 
[$(1+1)$D] conformal field theories (CFTs)
\cite{alexander_1984}.
Consequently, 
various lattice models with distinct microscopic details 
and high-energy spectra can flow to the same IR CFT.
Their low-energy states and operators share the same conformal operator content, 
organized into representations of the Virasoro algebra.
The resulting conformal data, 
including the central charge and scaling dimensions, 
determine the universal critical exponents and long-range correlations.

The universal content of a critical point is not 
confined to the fixed point itself.
A CFT together with a choice of relevant perturbation(s) may also
exhibit a universal off-critical scaling theory in the proximate regime.
Although the perturbation(s) modifies the IR physics and often open a gap, 
lattice systems with different UV spectra can still 
share the same IR massive scaling theory once non-universal scaling factors are fixed.
Along certain perturbative directions in $(1+1)$D, 
infinitely many local integrals of motion may persist, 
rendering the theory integrable and imposing strong constraints on its particle spectrum and factorized scattering~\cite{zam_1989,zam_1990}.
A celebrated example is the magnetic perturbation of the Ising CFT, 
which yields a massive integrable field theory with eight particles whose exact mass spectrum and scattering amplitudes are governed by the exceptional $E_8$ Lie algebra~\cite{delfino_1995,zam_1989,zam_1990}.
A particularly complete experimental realization of this spectrum was reported in the quasi-1D antiferromagnet BaCo$_2$V$_2$O$_8$ (BCVO) under a transverse field near its putative 1D quantum critical point
\cite{zhe_2018,yi_2019,haiyuan_2020,zhao,xiao_2021,zou,jiahao_review,xiao_2023}.
At the level of the effective lattice Hamiltonian, 
its low-energy magnetism is described by a spin-$1/2$ Heisenberg-Ising chain with four-site-periodic terms and a transverse field, 
substantially more complex than the canonical transverse-field Ising chain
\cite{haiyuan_2020,zou,jiahao_review,xiao_2023}.
Nevertheless, its critical scaling limit is governed by the Ising CFT, 
while the ordered background generated by weak interchain coupling acts as the effective longitudinal field that realizes the magnetic perturbation.
The resulting $E_8$ excitation spectrum, 
including all eight single-particles, 
was observed by inelastic neutron scattering, 
illustrating how universal spectral structures can emerge from microscopically complex lattice systems.

\begin{figure*}
    \centering
    \includegraphics[width=0.95\linewidth]{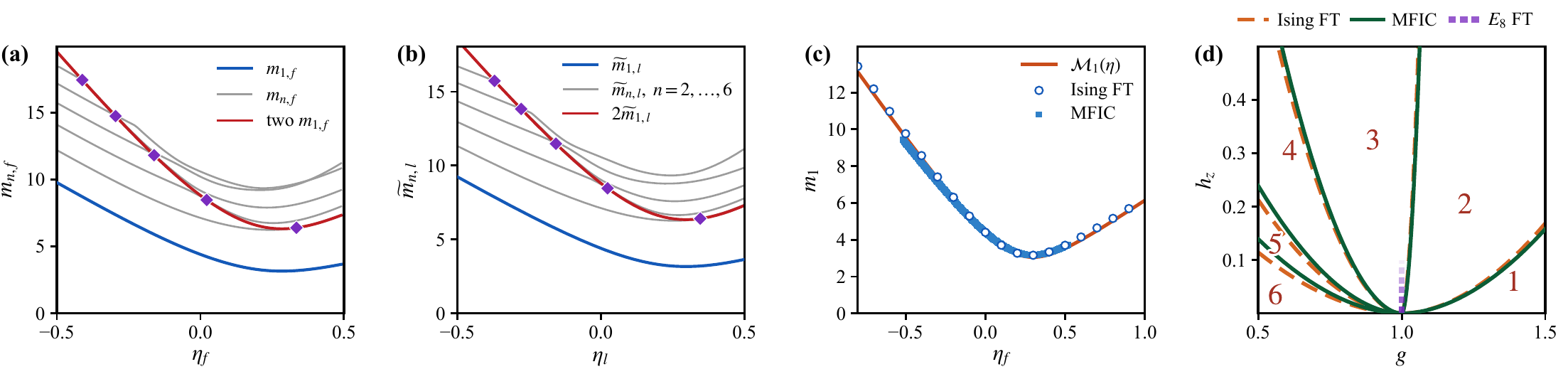}
    \caption{
Universal meson mass trajectories and stable-meson crossover windows.
(a) The six lowest meson masses of the Ising field theory as functions of $\eta_f=-\lambda/|h|^{8/15}$.
The field-theory data are calculated at $h=1$ by scanning $\eta_f$ from $-0.5$ to $0.5$ in steps of $0.01$.
The blue curve denotes $m_{1,f}$,
the gray curves denote $m_{n,f}$ with $n=2,\ldots,6$,
and the red curve denotes the lightest two-particle threshold $2m_{1,f}$.
The purple diamonds mark the crossings between the higher masses and $2m_{1,f}$.
(b) The corresponding reduced MFIC masses $\widetilde{m}_{n,l}$ as functions of $\eta_l$.
The MFIC calculation uses $h_z=0.018$ and is parameterized by
$\bar{\eta}_l=(g-g_c)/h_z^{8/15}$,
with $g=g_c+\bar{\eta}_l h_z^{8/15}$, $g_c=1$, and
$\eta_l=\mathcal{C}_f\bar{\eta}_l/(\pi\mathcal{C}_l)$.
(c) Comparison of the lightest-meson mass trajectories.
The orange curve is the field-theory analytical data generated from Eq.~(\ref{Eq:m1trajectory}).
The blue open circles are the reduced IFT results $m_{1,f}/|h|^{8/15}$,
and the filled squares are the reduced MFIC results $\widetilde{m}_{1,l}$ plotted at $\eta_f=\eta_l$.
(d) Stable-meson-count crossover boundaries in the $(g,h_z)$ plane.
The orange dashed curves are obtained from the IFT threshold crossings after converting to the lattice variables,
whereas the green solid curves are obtained directly from the MFIC spectrum.
The purple dotted line marks $g=g_c$, the $E_8$ scaling direction.
The numbers indicate the number of stable meson branches below the lightest two-particle threshold.
}
    \label{fig:1}
\end{figure*}

In this work,
we systematically investigate the universal scaling of meson spectra near $(1+1)$D Ising criticality.
Starting from the Ising field theory (IFT),
we use the truncated conformal space approach (TCSA) to verify numerically that the trajectory of the lightest meson mass $m_1$ is described by a continuous analytic scaling function of a single dimensionless parameter.
We then calculate the evolution of the meson spectrum in the near-critical mixed-field Ising chain (MFIC) using Hamiltonian truncation in the free-fermion basis.
After independently fixing the model-dependent normalizations of the energy-density and magnetic perturbations,
we find without further fitting that the reduced $m_1$ trajectory of the MFIC collapses onto the field-theory scaling function,
while the stable-meson-count crossover boundaries cut by the $2m_1$ threshold closely follow the field-theory predictions.

To connect the mass relations with the spectra,
we calculate the energy-density dynamical structure factor (DSF) of the IFT 
within TCSA and the transverse-spin DSF of the MFIC using the infinite time-evolving block decimation (iTEBD) method.
At representative values of the corresponding reduced scaling parameters,
the results exhibit the predicted numbers of isolated meson branches below the lightest two-particle continuum.
These results suggest that the universal scaling structure of the Ising CFT extends into the nearby confining regime and organizes its low-energy meson spectrum.
Accordingly,
we formulate a general framework for predicting the lightest-meson mass trajectory and stable-meson-count crossover windows in lattice realizations of $(1+1)$D Ising criticality from independently calibrated thermal and magnetic scaling factors.

As a concrete demonstration,
we apply this framework to a class of four-periodic spin-$1/2$ Heisenberg-Ising chains under transverse fields.
This Hamiltonian class contains a family of quantum Ising critical points
and includes effective spin models for BCVO and SrCo$_2$V$_2$O$_8$ (SCVO).
Taking BCVO as a representative material realization,
we independently determine its thermal and magnetic scaling factors using tensor-network calculations
and find that the reduced lightest-meson masses along the self-consistent field trajectory follow the IFT scaling function.
We further calculate the field-dependent transverse DSFs with iTEBD.
Across $1$--$9$ T,
the resolved branches exhibit the predicted succession of windows containing four, three, and two stable mesons below the $2m_1$ threshold.
Together,
these results predict the full field evolution of the low-energy meson spectrum in the 3D ordered region of BCVO
and provide a consistent interpretation of the spectroscopic features reported in Ref.~\cite{zhao}.

\textit{IFT and meson spectra.}---
Quantum critical points that belong to the $(1+1)$D Ising universality class are described by the $c=1/2$ CFT.
Near the CFT, we can generally define the IFT as
\begin{equation}
    \mathcal{H}_{\mathrm{IFT}}
    =
    \mathcal{H}_{c=1/2}
    -\lambda\int dx~\varepsilon(x)
    -h\int dx~\sigma(x).
    \label{Eq:isingFT}
\end{equation}
Here, $\mathcal{H}_{c=1/2}$ is the Hamiltonian of the Ising CFT,
while $\sigma(x)$ and $\varepsilon(x)$ are its two nontrivial primary fields,
known as the spin-density (magnetic) and energy-density fields, respectively.
If only a negative energy-density perturbation is present, namely $h=0$ and $\lambda<0$,
Eq.~(\ref{Eq:isingFT}) describes free massive Majorana fermions with the two-particle S-matrix $S=-1$.
The corresponding mass gap is then $\Delta_f=2\pi|\lambda|$.
Hereafter the subscript $f$ denotes field-theory quantities.
If only the magnetic perturbation is present, 
namely $\lambda=0$,
the theory becomes the integrable $E_8$ field theory,
which contains eight particles with mass ratios and S-matrices being governed by the exceptional $E_8$ Lie algebra~\cite{zam_1989,xiao_2021,delfino_1995}.
The mass of the lightest $E_8$ particle is given by $m_{1,f}=\mathcal{C}_{f}|h|^{8/15}$
with $\mathcal{C}_{f}\approx 4.40491$~\cite{fateev}.

When both perturbations are present, 
the field theory is generally non-integrable.
Nevertheless, 
its low-energy excitation spectrum can still be obtained numerically using TCSA~\cite{zam_1990,zam2003,TCSA}.
We then define a single scaling parameter as $\eta_f=-\lambda/|h|^{8/15}$.
Stable meson excitations remain in the whole parameter region due to the confinement induced by the magnetic perturbation.
Under a finite longitudinal field $h\neq0$, 
with $\eta_f\rightarrow+\infty$, 
the meson spectra approach the free-Majorana limit,
whereas with $\eta_f\rightarrow-\infty$, 
it gives rise to an infinite number of mesons
\cite{barry_1978,pedro_2006,alexander_2013,yunjing_2025,haolan_2024}.
For finite $\eta_f$ away from the integrable $E_8$ point,
we determine the number of stable mesons from the lightest two-particle threshold $2m_{1,f}$.
Meson branches below the threshold are identified as stable mesons with long lifetimes.
Furthermore,
the lightest-meson mass is a continuous function of $\eta_f$
and can be written as
$m_{1,f}=|h|^{8/15}\mathcal{M}_1(\eta_f)$.
Introducing $\zeta=2\pi\eta_f$,
the scaling function has the following analytical
representation~
\footnote{%
For the detailed analytical calculation, $\theta=11\pi/15$, $C_1\simeq1.295$. 
$\Delta_1(y)$ is the discontinuity across the Yang--Lee cut starting at $Y_0\approx2.4296$, and for $y\geq Y_0$ we set $T_0=Y_0^{-5/4}$. The discontinuity is represented by the seven-term interpolation of Ref.~\cite{haolan_2024},
$\Delta_1(y)=y\sum_{j=1}^{7} q_j\,\sin(\pi\alpha_j)e^{i\pi\alpha_j}\,\bigl(T_0-y^{-5/4}\bigr)^{\alpha_j}$,
with $\{\alpha_1,\ldots,\alpha_4\}=\{\tfrac{5}{12},\tfrac{5}{4},\tfrac{17}{12},\tfrac{25}{12}\}$ and $\{\alpha_5,\alpha_6,\alpha_7\}=\{\tfrac{9}{4},\tfrac{29}{12},\tfrac{11}{4}\}$. The first three real coefficients are
$q_1=\frac{\protect\widetilde{b}_0}{Y_0}\bigl(\tfrac{4}{5}Y_0^{9/4}\bigr)^{5/12}$,
$q_2=\frac{\protect\widetilde{c}_0}{Y_0}\bigl(\tfrac{4}{5}Y_0^{9/4}\bigr)^{5/4}$, and
$q_3=\frac{\protect\widetilde{b}_1}{Y_0}\bigl(\tfrac{4}{5}Y_0^{9/4}\bigr)^{17/12}-\frac{17q_1}{40T_0}$. The remaining four coefficients are determined by
$\sum_{j=1}^{7} q_j\,\sin(\pi\alpha_j)e^{i\pi\alpha_j}\,T_0^{\alpha_j}=E_1(0)$ and
$-\sum_{j=1}^{7} \alpha_j q_j\,\sin(\pi\alpha_j)e^{i\pi\alpha_j}\,T_0^{\alpha_j-1}=E_1'(0)$,
where $E_1(t)=t^{4/5}\Delta_1(t^{-4/5})$. Taking the real and imaginary parts gives four linear equations for $q_4,\ldots,q_7$. The numerical inputs are $\protect\widetilde{b}_0=3.0754$, $\protect\widetilde{c}_0=-0.9412$, $\protect\widetilde{b}_1=0.8932$, $E_1(0)=1.50\,i$, and $E_1'(0)=-0.5844-2.1810\,i$.
},
\begin{equation}
\begin{aligned}
\mathcal{M}_1&(\eta_f)
={}
\mathcal{C}_{f}
+C_1\zeta
\\
&+\frac{2\zeta^2}{\pi}
\int_{Y_0}^{\infty}\frac{dy}{y^2}
\frac{
y~\mathrm{Re}\left[e^{-i\theta}\Delta_1(y)\right]
-\zeta~\mathrm{Re}\left[\Delta_1(y)\right]
}{
y^2-2\zeta y\cos\theta+\zeta^2
}.
\end{aligned}
\label{Eq:m1trajectory}
\end{equation}

Such an IFT and its meson spectrum can be realized in various lattice systems,
since the associated low-energy structure is essentially an IR property that can be shared by different UV completions.
The simplest lattice model that realizes the IFT is the near-critical MFIC,
namely a near-critical transverse-field Ising chain with an additional weak longitudinal field.
The Hamiltonian reads
\begin{equation}
    \mathcal{H}_{\text{MFIC}}
    =
    -J\sum_i
    \left(
    \sigma_i^z\sigma_{i+1}^z
    +g\sigma_i^x
    +h_z\sigma_i^z
    \right).
    \label{Eq:TFIClat}
\end{equation}
Here, $\sigma_i^\alpha$ with $\alpha=x,y,z$ are the local Pauli matrices at site $i$,
and $J>0$ is the ferromagnetic coupling.
We take $J=1$ as the energy unit,
consider the thermodynamic limit,
and focus on $g$ near the critical value $g_c=1$.
Since the spectrum is invariant under $h_z\rightarrow-h_z$,
we restrict ourselves to $h_z>0$ and consider the limit $h_z\rightarrow0^+$ without loss of generality.
At $h_z=0$ and $g\neq g_c$,
the gap follows $\Delta_l=2|g-g_c|$,
and the low-energy excitations are free massive Majorana fermions.
Hereafter $l$ denotes lattice quantities for MFIC.
At $g=g_c$ and $h_z=0$,
the gap closes in the thermodynamic limit and the low-energy theory is the $c=1/2$ CFT,
whose conformal data can be extracted from the low-energy spectrum~\cite{yijian_2018,wei_2023}.
At $g=g_c$ and $h_z\rightarrow0^+$,
the model enters the $E_8$ scaling regime,
where the low-energy spectrum is described by the $E_8$ field theory.
The lightest mass then follows the scaling behavior $m_{1,l}=\mathcal{C}_{l}|h_z|^{8/15}$,
with $\mathcal{C}_{l}\approx 5.4154$ being the microscopic lattice constant~\cite{fateev,xiao_2024time,xiao_2025}.
More generally,
for $g\neq g_c$ within the near-critical confining regime and $h_z\rightarrow0^+$,
the low-energy spectrum is dominated by a sequence of confinement-induced meson excitations as discussed for the IFT meson spectra.

Using the Hamiltonian truncation method for both the IFT~\cite{TCSA} and the MFIC,
we obtain the meson mass spectra of the two models.
For each value of the corresponding scaling parameter,
we retain the lowest 20 eigenvalues and extract the six lowest excitation gaps,
denoted by $m_1,\ldots,m_6$,
at zero momentum.
Comparing these gaps with the lightest two-particle threshold $2m_1$,
we identify the number of meson branches that remain stable away from the integrable $E_8$ point in the IFT [Fig.~\ref{fig:1}(a)] and the MFIC [Fig.~\ref{fig:1}(b)].
For the IFT,
we employ TCSA~\cite{TCSA}.
We work in the zero-momentum sector,
and impose the truncation cutoff with corresponding conformal basis contains 4076 states.
The numerically obtained reduced trajectory $m_{1,f}/|h|^{8/15}$ as a function of $\eta_f$ agrees very well with the analytical result in Eq.~(\ref{Eq:m1trajectory}) [Fig.~\ref{fig:1}(c)].
For the MFIC,
we consider a periodic chain of length $L=100$ in the zero total-momentum sector,
rebuild the free-fermion basis at each transverse field $g$,
and retain the lowest $N_{\mathrm{cut}}=8192$ basis states.
Furthermore,
we introduce a common energy scale and define the rescaled scaling parameter and meson masses as
\begin{equation}
    E_l=\frac{\mathcal{C}_l}{\mathcal{C}_f}h_z^{8/15},
    ~
    \eta_l=\frac{g-g_c}{\pi E_l},
    ~
    \widetilde{m}_{n,l}=\frac{m_{n,l}}{E_l}.
    \label{Eq:MFICRescaling}
\end{equation}
With the rescaling,
we find that the MFIC trajectory $\widetilde{m}_{1,l}(\eta_l)$ agrees,
within numerical accuracy,
with both the analytical result in Eq.~(\ref{Eq:m1trajectory}) and the numerical IFT trajectory [Fig.~\ref{fig:1}(c)].
Based on the same threshold criterion,
we further determine the stable-meson-count crossover boundaries for the IFT and the MFIC.
After mapping the field-theory crossing positions into the lattice parameter plane $(g,h_z)$ using the same scaling relation,
we find that they nearly coincide with the boundaries obtained directly from the MFIC over broad parameter range shown in Fig.~\ref{fig:1}(d).
\begin{figure*}
    \centering
    \includegraphics[width=0.98\linewidth]{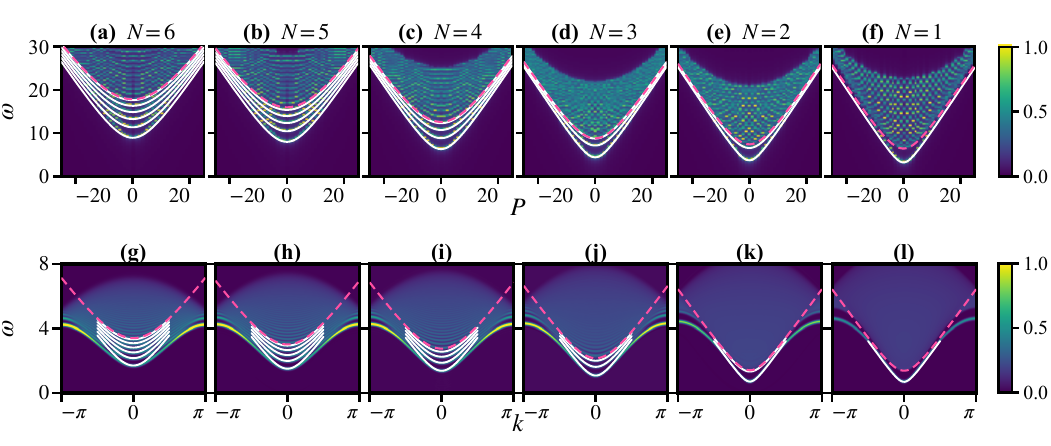}
    \caption{
Dynamical signatures of the meson crossover windows.
(a)--(f) Normalized energy-density DSFs
$S^{\varepsilon\varepsilon}(\omega,P)$
of the IFT at
$\eta_f=-0.43,-0.35,-0.20,0,0.10$, and $0.35$,
respectively.
(g)--(l) Corresponding normalized DSFs
$S^{xx}(\omega,k)$
of the MFIC obtained by iTEBD.
For these panels,
the longitudinal field is $h_z=0.036$,
and the transverse fields
$g=0.74,0.80,0.84,0.94,1.12$, and $1.26$
correspond to the rescaled parameters
$\eta_l=-0.40,-0.30,-0.24,-0.091,0.18$, and $0.40$,
respectively.
Each set of panels follows the same sequence of threshold-based meson counts
$N=6,5,\cdots,1$.
The white solid curves show the first $N$ single-meson relativistic energy-momentum dispersion guides.
The magenta dashed curves denote the lower boundary of the two-lightest-meson continuum.
}
\label{fig:2}
\end{figure*}

To identify observable signatures of the meson spectra,
a general route is to calculate the DSF,
defined as
\begin{equation}
    S^{\mathcal{O}\mathcal{O}}(\omega,k)
    =
    \frac{1}{L}
    \sum_{j}
    \int_{-\infty}^{\infty}dt~
    e^{i\omega t-ikj}
    \left\langle
    \delta\mathcal{O}_j(t)\delta\mathcal{O}_0(0)
    \right\rangle .
    \label{Eq:DSF}
\end{equation}
Here,
$\delta\mathcal{O}_j(t)=\mathcal{O}_j(t)-\langle\mathcal{O}_j\rangle$.
For the IFT,
we choose $\mathcal{O}=\varepsilon$ and calculate the DSF using TCSA.
The integer TCSA momentum quantum numbers $P=-40,-39,\ldots,40$ are included in the final results.
The discrete spectral peaks are represented by a Lorentzian broadening with $a=0.08$,
and the DSF is evaluated on the frequency grid $\omega=0.1,0.2,\ldots,30.0$.
The results are shown in Fig.~\ref{fig:2}(a)--(f).
For the MFIC,
we choose $\mathcal{O}=\sigma^x$ and calculate the DSF directly in the thermodynamic limit using the iTEBD method.
Near the critical point,
the transverse-spin operator has the expansion
$\sigma_i^x=A_0\mathbf{1}+A_{\varepsilon}\varepsilon(x)+\cdots$,
so its connected response probes the same energy-density channel as the field-theory calculation.
We use bond dimension $D=32$,
time step $\delta t=0.4$,
and maximum evolution time $t_{\max}=400$ for the iTEBD calculations.
The results are shown in Fig.~\ref{fig:2}(g)--(l).
In both models,
the bright low-energy ridges follow the relativistic meson dispersions determined from the zero-momentum masses in Fig.~\ref{fig:1}.
As the system passes through the successive threshold windows,
the number of isolated branches below the lightest two-particle continuum decreases from six to one.
Although the detailed spectral intensities and high-energy continua remain model dependent,
the low-energy branch counting follows the same meson-spectrum organization in the IFT and the MFIC.

\textit{Universal scaling of the meson spectra.}---
The agreement between the MFIC and IFT suggests a universal organization of the meson spectra near $(1+1)$D Ising criticality 
with the corresponding microscopic parameters rescaled.
It indicates that the low-energy meson spectra in the IR limit is governed by the massive scaling theory emanating from the Ising fixed point.
Thus the universal scaling structure of Ising criticality extends from critical point to the organization of the nearby confining spectrum.
We thus further propose the following framework for understanding the low-energy excitations of Ising magnets.
For any lattice system with a $(1+1)$D quantum Ising critical point $g_{c,u}$,
we denote by $g_u$ and $h_u$ the microscopic parameters coupling to the energy-density and spin-density directions, respectively.
Hereafter the subscript $u$ labels a generic microscopic realization of $(1+1)$D Ising criticality.
The corresponding nonuniversal scaling factors can be determined independently by switching on one perturbation at a time.
Choosing $g_u-g_{c,u}>0$ on the disordered side,
we define the signed thermal scale and the magnetic mass scale as
\begin{equation}
    \Delta_u^{\varepsilon}=A_u(g_u-g_{c,u}),
    ~
    \Delta_u^{\sigma}=\mathcal{C}_u|h_u|^{8/15}.
    \label{Eq:MicroscopicScales}
\end{equation}
Here, $|\Delta_u^{\varepsilon}|$ is the excitation gap at $h_u=0$,
with the sign of $\Delta_u^{\varepsilon}$ distinguishing the two sides of the critical point,
while $\Delta_u^{\sigma}$ is the lightest $E_8$ particle's mass at $g_u=g_{c,u}$.
Using the field-theory mass amplitude $\mathcal{C}_f$,
we define the rescaled scaling parameter and reduced meson masses as
\begin{equation}
\begin{aligned}
    \eta_u
    =
    \frac{A_u\mathcal{C}_f}{2\pi\mathcal{C}_u}
    \frac{g_u-g_{c,u}}{|h_u|^{8/15}},~
    \widetilde{m}_{n,u}
    =
    \frac{\mathcal{C}_f}{\mathcal{C}_u}
    \frac{m_{n,u}}{|h_u|^{8/15}}.
\end{aligned}
    \label{Eq:UniversalMesonRescaling}
\end{equation}
For the IFT,
$\Delta_f^{\varepsilon}=-2\pi\lambda$ with $\Delta_f=|\Delta_f^{\varepsilon}|$, and
$\Delta_f^{\sigma}=\mathcal{C}_f|h|^{8/15}$,
so Eq.~(\ref{Eq:UniversalMesonRescaling}) reproduces
$\eta_f=-\lambda/|h|^{8/15}$.
For the MFIC,
$A_l=2$ with $\Delta_l=|\Delta_l^{\varepsilon}|$, and Eq.~(\ref{Eq:UniversalMesonRescaling}) reduces to the rescaling in Eq.~(\ref{Eq:MFICRescaling}).
Once $g_{c,u}$, $A_u$, and $\mathcal{C}_u$ are determined independently,
no additional fitting parameter enters the comparison.
Within the scaling regime,
the reduced mass trajectories $\widetilde{m}_{n,u}(\eta_u)$ should approach the Ising-field-theory predictions.
The crossings with the reduced two-particle threshold $2\widetilde{m}_{1,u}$ then determine the corresponding stable-meson-count crossover windows.

While we have benchmarks based on the simplest MFIC Hamiltonian as the above results,
we try to show the above framework really works universally for the general Ising criticality.
To test this,
in the following we consider a class of four-periodic spin-$1/2$ Heisenberg-Ising chains described by
\begin{equation}
\begin{aligned}
\frac{\mathcal{H}_{\mathrm{HI}}}{J_z}
={}&\sum_i\left[
s_i^zs_{i+1}^z
+\epsilon\left(s_i^xs_{i+1}^x+s_i^ys_{i+1}^y\right)
\right]
\\
&-H_x\sum_i\left[
s_i^x+C_y(-1)^is_i^y
+C_z\cos\!\left(\frac{\pi i}{2}-\phi\right)s_i^z
\right]
\\
&-H_z\sum_i(-1)^is_i^z .
\end{aligned}
\label{Eq:FourPeriodic}
\end{equation}
Here, $s_i^\alpha=\sigma_i^\alpha/2$ with taking $\hbar=1$,
and all energies and fields are expressed in units of the Ising exchange $J_z$.
The parameter $\epsilon$ is the transverse exchange.
$C_y$ and $C_z$ describe the induced staggered transverse field
and the four-periodic longitudinal field generated by the helical local $g$ tensor, respectively.
The phase is fixed at $\phi=\pi/4$,
and the last term is an additional staggered longitudinal field.
This Hamiltonian is known to describe the effective spin Hamiltonians for both the quais-1D anti-ferromagnets BCVO and SCVO within their 3D ordered phase~\cite{zhe_2018,yi_2019,zhao,zou,haiyuan_2020,jiahao_review,xiao_2023}.
The parameters $(\epsilon,C_y,C_z)$ serve as material-dependent microscopic parameters for both the two materials.
For BCVO,
we take $(\epsilon,C_y,C_z)=(0.47,0.40,0.14)$,
whereas for SCVO we take $(0.4662,0.29,0.14)$
\cite{zhe_2018,yi_2019,haiyuan_2020,zhao,xiao_2021,zou,jiahao_review,xiao_2023}.
At $H_z=0$,
each parameter set exhibits an Ising quantum critical point at a material-dependent critical field.

\begin{figure}
    \centering
    \includegraphics[width=\linewidth]{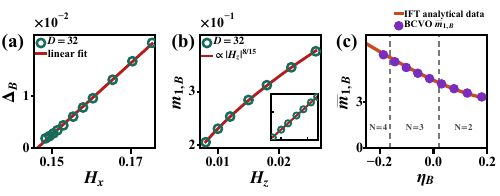}
    \caption{
Calibration of the BCVO scaling variables and lightest-meson trajectory.
(a) Lowest excitation gap at $H_z=0$ obtained from the $D=32$ VUMPS calculation.
The red line is a linear fit in $H_x$.
(b) Lightest mass at the independently calibrated critical field $H_x=H_{c,B}=0.1454$ as a function of $H_z$.
The red line is a fit with exponent $8/15$,
and the inset shows the same data on logarithmic scales.
(c) Reduced lightest-meson masses along the self-consistent BCVO field path from $1$ to $9$ T.
The orange curve is the field-theory analytical data generated from Eq.~(\ref{Eq:m1trajectory}), evaluated at $\eta_f=\eta_B$,
the purple circles are the BCVO results,
and the gray dashed lines are the mapped threshold crossings.
The labels indicate the stable-meson count in each interval.
}
    \label{fig:3}
\end{figure}

We use BCVO as a representative material realization and determine its two microscopic scaling factors with variational uniform matrix product state (VUMPS) algorithm~\cite{valentin_2018}.
For BCVO, the variables in the universal framework are identified as $g_B=H_x$, $h_B=H_z$, and $g_{c,B}=H_{c,B}$.
The disordered-side thermal gap and the lightest mass along the magnetic direction are described by
\begin{equation}
    \Delta_B
    =A_B\left(H_x-H_{c,B}\right),
    ~
    m_{1,B}
    =\mathcal{C}_B |H_z|^{8/15}.
    \label{Eq:BCVOCalibration}
\end{equation}
The subscript $B$ denotes BCVO-related quantities.
The fits give
$H_{c,B}=0.1454$, $A_B=0.6882$, and
$\mathcal{C}_B=2.5470$,
as shown in Fig.~\ref{fig:3}(a) and (b).
Using these coefficients, the BCVO reduction reads
\begin{equation}
\begin{aligned}
    \eta_B=
    \frac{A_B\mathcal{C}_f}{2\pi\mathcal{C}_B}
    \frac{H_x-H_{c,B}}{|H_z|^{8/15}},~
    \widetilde{m}_{1,B}=
    \frac{\mathcal{C}_f}{\mathcal{C}_B}
    \frac{m_{1,B}}{|H_z|^{8/15}}.
\end{aligned}
\label{Eq:BCVORescaling}
\end{equation}
All quantities entering Eq.~(\ref{Eq:BCVORescaling}) are fixed by the two independent scans.

In the 3D ordered phase of BCVO,
weak interchain coupling induces the staggered longitudinal field.
Within a self-consistent chain mean-field treatment,
its magnitude satisfies
$H_z=\lambda_B|m_s|$,
where $m_s=\frac{1}{4}\sum_{a=1}^{4}(-1)^a\langle s_a^z\rangle$
and $\lambda_B\approx 0.048$
\cite{zhe_2018,yi_2019,haiyuan_2020,zhao,xiao_2021,zou,jiahao_review,xiao_2023}.
For the physical transverse field,
we use $H_x=0.0274B$ with $B$ the magnitude in Tesla.
Increasing $B$ therefore changes both relevant couplings and traces a path through the confining regime.
The finite $H_z$ maintains confinement and produces a meson spectrum throughout this path.
After applying Eq.~(\ref{Eq:BCVORescaling}),
the reduced lightest masses extracted along the $1$-$9$ T path agree with the field-theory analytical data generated from Eq.~(\ref{Eq:m1trajectory}) after identifying $\eta_f=\eta_B$ [Fig.~\ref{fig:3}(c)].
The IFT threshold crossings mapped to $\eta_B$ then predict the stable-meson count along this path.

\begin{figure}
    \centering
    \includegraphics[width=\linewidth]{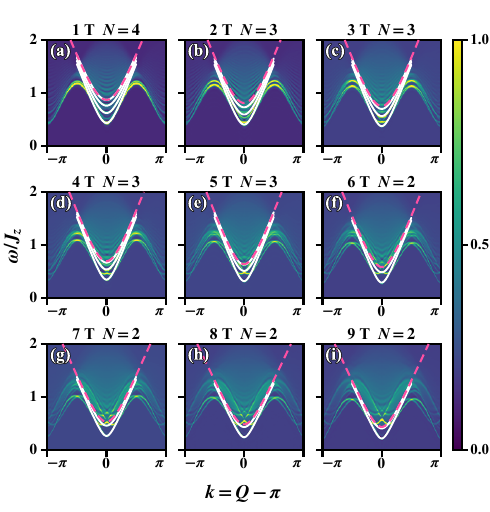}
    \caption{
Field evolution of the BCVO meson spectrum.
(a)--(i) Normalized combined transverse DSFs
$S^{xx}(\omega,Q)+S^{yy}(\omega,Q)$
for $B=1,2,\ldots,9$ T,
using the self-consistent $H_z$ at each field.
The momentum is measured from the antiferromagnetic wave vector,
$k=Q-\pi$.
The white solid curves show the first $N$ one-meson dispersion guides obtained from the mapped IFT spectrum.
The magenta dashed curve is the corresponding two-lightest-meson threshold guide.
The panel headings give the threshold-based stable-meson count $N$.
}
    \label{fig:4}
\end{figure}

To resolve these modes,
we calculate the DSFs $S^{xx}(\omega,Q)$ and $S^{yy}(\omega,Q)$ by setting $\mathcal{O}=s^x$ and $s^y$ in Eq.~(\ref{Eq:DSF}), where the Fourier momentum is denoted by $Q$ in this section.
In the unfolded momentum representation,
$S^{yy}(\omega,Q=\pi)$ directly resolves the meson spectrum.
The four-site unit cell produces zone-folded shadows of these modes at shifted momenta.
The iTEBD calculation reconstructs the correlations on the physical lattice and retains the full Brillouin zone,
so the exact-$Q=\pi$ response remains identifiable in the presence of the shadow branches.
For comparison to the inelastic neutron scattering experiments,
we combine the two transverse channels.

We show the field-dependent spectra follow this prediction in Fig.~\ref{fig:4}.
At $k=0$,
four primary branches lie below $2m_1$ at $1$ T,
three remain below the threshold from $2$ to $5$ T,
and two remain from $6$ to $9$ T.
Throughout these intervals,
the modes resolved in the $Q=\pi$ channel follow the predicted dispersions,
and their number agrees with the threshold criterion.
The same assignment provides a consistent interpretation of the field-dependent spectral features reported in the Appendix of Ref.~\cite{zhao}.
The agreement thus extends the quantum Ising criticality to the universal meson scaling 
in the quasi-1D magnets effectively described by
the quantum Ising Hamiltonian.

\textit{Discussion and Conclusion.}---
In this work,
we systematically study the meson spectra near $(1+1)$D Ising criticality
and establish their universal organization.
For the IFT,
the meson masses across the relevant deformations are obtained with TCSA,
and we show that the lightest mass follows the analytical trajectory controlled by a single scaling parameter.
The crossings with the $2m_1$ threshold are then used
to determine the successive stable-meson-count crossover windows.
The MFIC is next considered as the simplest lattice realization of the IFT.
Its meson spectrum is calculated using Hamiltonian truncation,
and the microscopic thermal and magnetic scaling factors are determined independently.
After the corresponding rescaling,
we show that both the lightest-meson trajectory and the crossover boundaries agree with the field-theory results.
The DSFs of the two models are also calculated,
and we show that their isolated low-energy branches exhibit the same sequence of stable-meson counts.
We thereby place the integrable $E_8$ spectrum within a continuous family of confining meson spectra
emanating from quantum Ising criticality.

Based on this correspondence,
we formulate a general framework for lattice models near quantum Ising criticality.
The critical point and the two scaling factors associated with the thermal and magnetic perturbations
are first determined
and then used to map the microscopic masses and couplings onto the field-theory scaling variables.
This procedure is applied to the four-periodic Heisenberg-Ising Hamiltonian describing BCVO.
The two scaling factors and the reduced lightest-meson masses
along the self-consistent field trajectory generated by the 3D ordered state
are obtained with VUMPS.
We show that these masses follow the IFT scaling function over the physical field range considered.
The DSFs are further calculated, which directly confirms the universal IFT predictions for the stable mesons in BCVO across a wide transverse field range.
We thus predict the field evolution of the low-energy BCVO meson spectrum.
The same threshold criterion provides a consistent interpretation
of the field-dependent spectral features reported in Ref.~\cite{zhao}.

We therefore establish a universal description of Ising meson spectra.
After the microscopic couplings are independently calibrated and rescaled,
we show that the meson spectra of different microscopic models follow the same universal scaling structure.
We further expect this construction to extend to other quantum critical universality classes
that admit integrable massive perturbations with well-defined quasiparticle spectra.
Starting from the exact spectrum along an integrable direction,
we expect that the calibration of the relevant microscopic couplings
and the tracking of the quasiparticle masses and multiparticle thresholds away from that direction
will reveal analogous universal trajectories and stability windows.
We thus provide a route for connecting integrable field theories,
microscopic many-body Hamiltonians,
and experimentally measured quasiparticle spectra near quantum criticality.

\textit{Acknowledgments.}--- We thank X. He for helpful discussions. 
J. Wu is sponsored by the National Natural Science Foundation of China Nos. 12274288, 12450004, 
the Innovation Program for Quantum Science and Technology Grant No. 2021ZD0301900,
and the Fundamental Research Funds for the Central
Universities. 
X. Wang is supported by the U.S. Department of Energy through Award Number DE-SC0023905.
The computation was performed on high-performance computing clusters supported by the Gordon and Betty Moore Foundation's EPiQS Initiative, Grant GBMF10436.
Part of the tensor network calculations were performed using the TeNPy Library~\cite{tenpy2024}.
The TCSA calculation is supported by the TCSA package in~\cite{TCSA}.
J.W. acknowledges the hospitality of Wilczek Quantum Center at Shanghai Institute for Advanced Studies of University of Science and Technology of China.

\bibliography{main}

\end{document}